\documentclass[preprint]{aastex}

\newcommand{\etal}{{\it et al.}}
\newcommand{\Halpha}{H$\alpha$}

\slugcomment{Accepted for publication in the {\it Astrophysical Journal}}

\shorttitle{SFRs IN EVOLVING GALAXY POPULATIONS}
\shortauthors{Afonso, Cram, \& Mobasher}

\begin{document}

\title{ON THE DETERMINATION OF STAR FORMATION RATES IN EVOLVING GALAXY
POPULATIONS}

\author{J. Afonso}
\affil{Blackett Laboratory, Imperial College, 
Prince Consort Rd, London SW7 2BW, UK}
\email{j.afonso@ic.ac.uk}

\author{L. Cram}
\affil{School of Physics, University of Sydney, Sydney NSW 2006, Australia}
\email{l.cram@physics.usyd.edu.au}

\and

\author{B. Mobasher}
\affil{Blackett Laboratory, Imperial College, Prince Consort Rd, London SW7 2BW, UK}
\email{b.mobasher@ic.ac.uk}

\begin{abstract}
The redshift dependence of the luminosity density in certain wavebands
(e.g. UV and \Halpha) can be used to infer the history of star
formation in the populations of galaxies producing this
luminosity. This history is a useful datum in studies of galaxy
evolution. It is therefore important to understand the errors that
attend the inference of star formation rate densities from luminosity
densities. This paper explores the self-consistency of star formation
rate diagnostics by reproducing commonly used observational procedures
in a model with known galaxy populations, evolutionary histories and
spectral emission properties. The study reveals a number of potential
sources of error in the diagnostic processes arising from the
differential evolution of different galaxy types. We argue that
multi-wavelength observations can help to reduce these errors.
\end{abstract}


\section{INTRODUCTION}

The stellar content and hence the spectral energy distribution (SED) of a
galaxy depends on many factors. Accurate predictions of galaxy SEDs require
sound theories of stellar evolution and stellar atmospheres, including
transient and extreme phases that remain difficult to model. In addition,
the evolution of a galaxy SED depends on (i) the initial mass function
(IMF) of stars and (ii) the history of the star formation rate (SFR). Parts
of the SED, such as the UV continuum and the Balmer lines, are sensitive to
the recent IMF and SFR, while other parts, such as the IR continuum,
reflect long-term averages. This opens the possibility of using
observations of SEDs to determine the current and past SFR in a particular
galaxy, and, by observing the SEDs of populations of galaxies over a range
of redshifts, the history of star formation in the universe.  To do this,
one needs a galaxy spectral synthesis model connecting SEDs with SFRs and
IMFs.  Early models of this kind were reviewed by \citet{Tin80}, and more
recent work is summarised by \citet{Lei96} and \citet{Sch99}.

The past decade or so has seen the acquisition of a rapidly growing
body of data on the distribution, over redshift, absolute luminosity,
and galaxy type, of those spectral properties of galaxies sensitive to
star formation rates.  Particularly notable have been new data on
rest-frame UV luminosities to redshifts $z \approx 4$
\citep{Lil96,Mad96, Cow97,Con97,Mad98,Pas98,Trey98,Cow99,Sul99},
emission line luminosities in \Halpha\ and [O{\sc ii}] $\lambda372$~nm
to $z \approx 1$ \citep{Gal95,Cow97,Tre98, Gla98}, and photometry
(with redshift estimates to $z \approx 1$) in the far-IR, sub-mm and
radio spectral bands \citep{Row97,Flo99,Hug98,Bla99, Cram98}. The data have
been interpreted by several of these authors using galaxy spectral
synthesis models, to yield estimates of the star formation rate as a
function of redshift.

Although it is widely recognised that there are numerous sources of
uncertainty in the process of inferring star formation rates from the
observable diagnostics, there have been few systematic, internally
consistent investigations of these uncertainties \citep[see also][]{Sch99}. 
In an attempt to bridge this gap, this paper uses a galaxy
spectral evolution code to test the self-consistency of common
diagnostic procedures.  We do this by comparing the known star
formation history of a model universe containing specified galaxy
populations with the star formation history that would be inferred by
applying commonly adopted diagnostic procedures.  Two questions are 
addressed: (1) are star formation rates derived from \Halpha\ and UV 
luminosities consistent with each other?, and (2) are star formation rates 
inferred from luminosity densities consistent with the true star formation 
rate in the model? It is important to stress that we do not aim to explore
the validity of any particular model of cosmic star formation history:
we are concerned here only with checking the internal consistency of
diagnostic procedures.

\section{THE MODEL AND ITS CALIBRATION}

We use the galaxy spectral evolution model {\sc pegase} \citep{Fio97}
and the galaxy population evolution model of Pozzetti, Bruzual \&
Zamorani (1996) to predict the evolution of the \Halpha\ and UV
luminosity densities in a ``model universe'' with a known star
formation history.  The key steps in our approach are (1) calculate
the evolution of the actual SFR density defined by the parameters
given in the model; (2) combine {\sc pegase} and the model universe to
predict the evolving luminosity densities; (3) use {\sc pegase} to
calibrate the SFR in terms of luminosity density using the methods
commonly applied to observations, and (4) combine the calibration and
the predicted luminosity density evolution to deduce the SFR history
for comparison with step (1).

\citet{Poz96} explored pure luminosity evolution (PLE) models based on
a mix of four galaxy types, E/S0, Sab-Sbc, Scd-Sdm and very Blue
(vB). The different types are denoted hereinafter by the parameter
$k$. The local luminosity function $\Phi_k(L)$ of each type in each
selected waveband is parametrised by the local space density
$\Phi_k^{\ast}$, characteristic luminosity $L^{\ast}_k$, and faint-end
slope $\alpha_k$. Each type also has a characteristic IMF $\Psi_k(M)$
and star formation rate history, $\dot{\rho}_k(t)$. A Scalo-type IMF
is used for the E/S0 and Sab-Sbc types, hereinafter called ``early'',
while a Salpeter-type IMF is used for the Scd-Sdm and vB types,
hereinafter called ``late''. For the E/S0 galaxies, Pozzetti \etal\
consider two models distinguished by different e-folding times
($\tau_1, \tau_2$) in their SFR. We adopt the $\tau_{2}$ model.

\citet{Poz96} constructed their model universe to match a number of
observational constraints, including the source count distribution in
several optical and IR photometric bands, the distribution of colours
as a function of apparent magnitude, and the distribution of redshifts
as a function of magnitude.  \citet{Poz96} exhibit a PLE model which,
in an $\Omega=0$ Friedmann cosmology, leads to acceptable agreement
with almost all of these constraints.  They also deduce that PLE
models in a flat ($\Omega=1$) cosmology cannot reproduce several
aspects of the data, and therefore we consider only the $\Omega=0$ and
$H_0=50$~km~s$^{-1}$~Mpc$^{-1}$ model.  

One constraint not used by Pozzetti \etal\ is the observed redshift
dependence of the luminosity density in certain wavebands.
Figure~\ref{fig:ldens} compares the UV luminosity density (${\cal
L}^{200}$ -- see below) of the model of Pozzetti \etal\ with the
observations of \citet{Cow99}. While there remain significant
uncertainties in the measured UV luminosity density
\citep[cf.][]{Lil96,Cow99,Sul99}, there is satisfactory agreement
between the prediction and recent measurements. The significance of
this will be amplified below.

Our application of {\sc pegase} takes place in two steps. First, for
galaxies of type $k$ we compute the time-dependent spectral emission
which follows the instantaneous formation of 1 M$_\odot$ of stars. We
use the evolutionary tracks of \citet{Bre93} supplemented to later
evolutionary phases and to lower masses as indicated in \cite{Fio97}.
We use the spectral stellar library described by \cite{Fio97}.  We
ignore extinction in the prediction of the UV luminosity, and assume
the number of ionizing photons to be 70\% of the Lyman continuum
photons. To ensure agreement with the evolutionary tracks, {\sc
pegase} has an upper limit of 120 M$_{\odot}$ for the chosen IMFs.
This leads to a minor inconsistency with the upper limit of 125
M$_{\odot}$ used by Pozzetti \etal, but this has no effect on our
conclusions.

In the second step we convolve the time-dependent spectral emission
with the star formation rate history $\dot{\rho}_k(t)$, to determine the 
evolution of $L^{\ast}_k$. Inserted into the luminosity function in the 
model of Pozzetti \etal\ this yields a prediction of the luminosity density
${\cal L}_k^p(t)$ produced in the (rest-frame) waveband $p$ at time
$t$ by the population $k$ undergoing star formation with a rate
density of $\dot{\rho}_k(t)$. The total SFR density is then clearly
\[
\dot{\rho}(t) = \sum_k \dot{\rho}_k(t),
\]
and the total luminosity density in waveband $p$ is
\[
{\cal L}^p(t) = \sum_k {\cal L}_k^p(t).
\]
For illustration it is convenient to define the ratio 
\[
R_k^p(t) = \dot{\rho}_k(t)/{\cal L}_k^p(t).
\]
It is also convenient to define a
global ratio for waveband $p$ as
\[
R^p(t) = \dot{\rho}(t)/{\cal L}^p(t).
\]
Our notation recognises that $R$ may be time dependent, and may depend
on the galaxy type $k$ and waveband $p$.

\begin{figure}
\plottwo{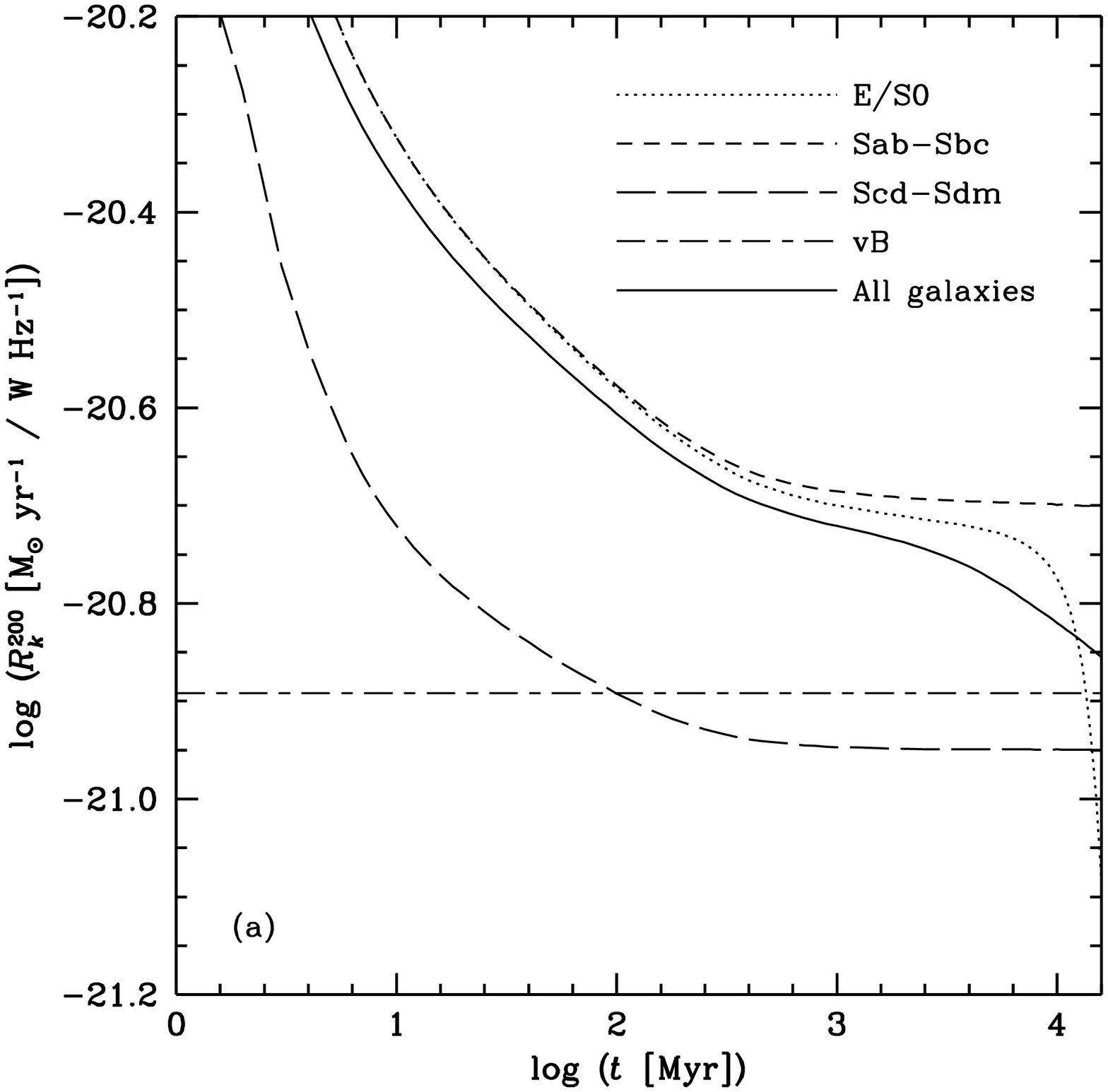}{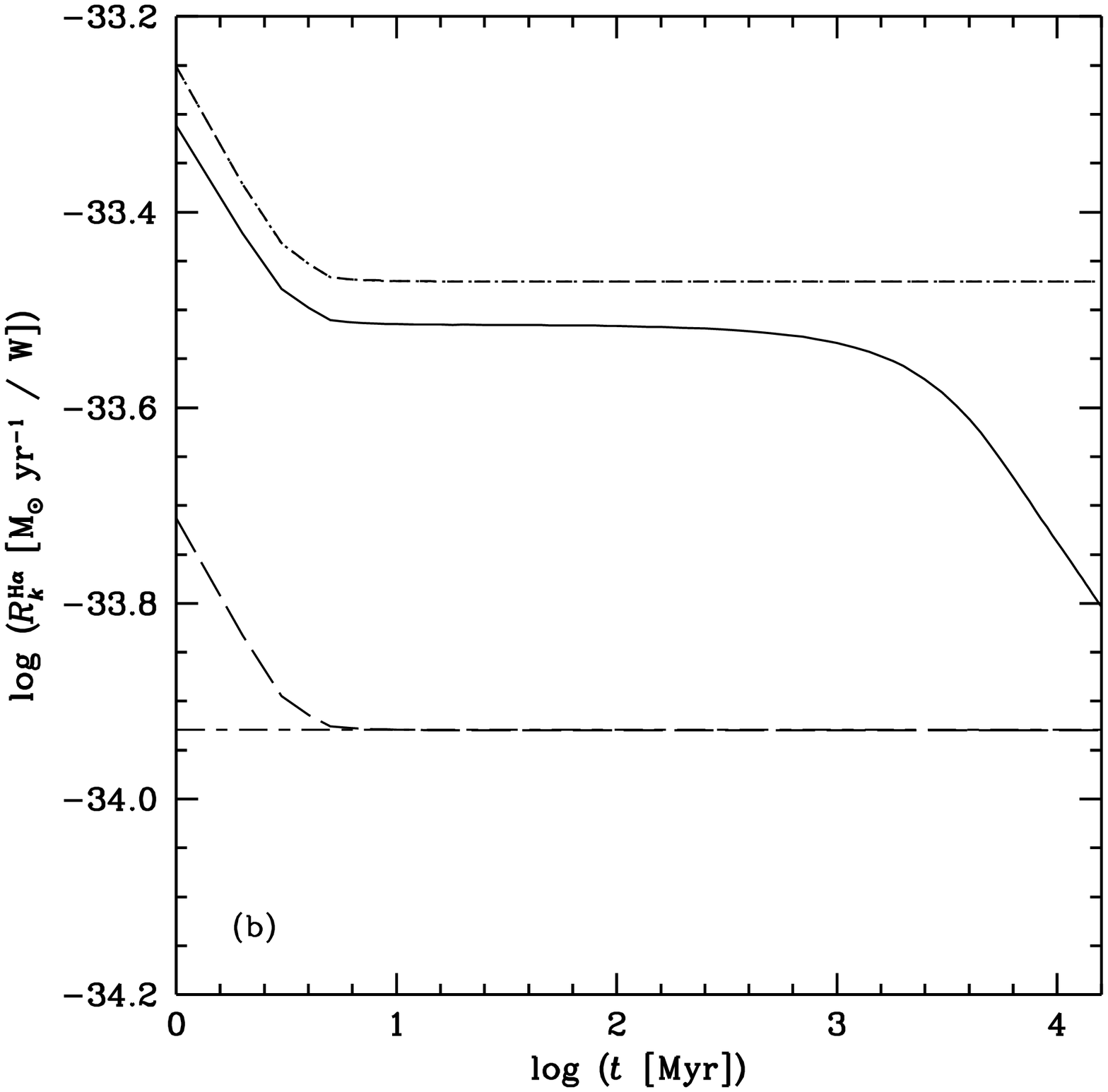}
\figcaption[Fig2a.eps,Fig2b.eps]{(a) The evolution with time of the
ratio of SFR density to luminosity density at 200~nm for each galaxy
type considered in this study. The solid line represents the evolution
of the global ratio of the total star formation rate density to
luminosity density, $R^{200}(t)$. (b) The same as (a), for
\Halpha. \label{fig:cevol}}
\end{figure}

Figures~\ref{fig:cevol}(a) and (b) exhibit the evolution with time of
$R_k^p(t)$ for each galaxy type and of the global ratio $R^p(t)$,
respectively for the 200~nm continuum and for \Halpha.  As previously
stressed by \citet{Ken83}, \citet{Sch99} and others, the ratio for
\Halpha\ in each galaxy type rapidly settles to a steady value,
reflecting the fact that \Halpha\ emission is completely controlled by
the short-lived, massive component of the IMF. The difference of
$\approx 0.5$ dex between the asymptotic ratios for the early and
late-type galaxies is due to the adoption of Scalo and Salpeter IMFs,
respectively, for the two types.

The evolution of $R_k^p(t)$ for the 200~nm continuum is quite
complex. For the E/S0 type, in which most star formation takes place
in the first 2 Gyr, the ratio displays a steady decline over $\approx
1$ Gyr, a plateau to $\approx 8$ Gyr, and a subsequent decline to the
present epoch. The initial decline arises from the rapid evolution of
the initial burst, while the later decline reflects the late stages of
evolution of relatively low mass stars at a time when few new stars
are being born. The extended plateaux in the Sab-Sbc and Scd-Sdm types
reflect the slower change in the star formation rate in these
populations, while the difference in asymptotic value is due to the
different IMFs. By definition, the vB component has a ratio equal to
that of the Scd-Sdm type at an age of 100 Myr.

To emulate observational procedures, we require calibration constants
$C^p$ that do not depend on time or galaxy type. These have been
estimated by previous workers using galaxy spectral synthesis models
over a range of star-forming histories and ages, and selecting a
``typical'' value \citep[e.g.][]{Ken94,Sch99}. We have conducted a
similar study using {\sc pegase}. As with previous derivations of
calibration factors, we find a significant sensitivity to the IMF, but
this is not the focus of our study. Accordingly, we explore
calibrations based on both the Salpeter (``late'') and Scalo
(``early'') models adopted by Pozzetti \etal\ The calibration
constants are listed in Table~\ref{tab:cs}.  We acknowledge that there
is inevitable uncertainty in the precise numerical values of the
calibration factors in our study, as there is in other studies, but
stress that this uncertainty has no effect on our conclusions.

\section{THE INFERRED STAR FORMATION RATE AND ITS EVOLUTION}

\begin{figure}
\plotone{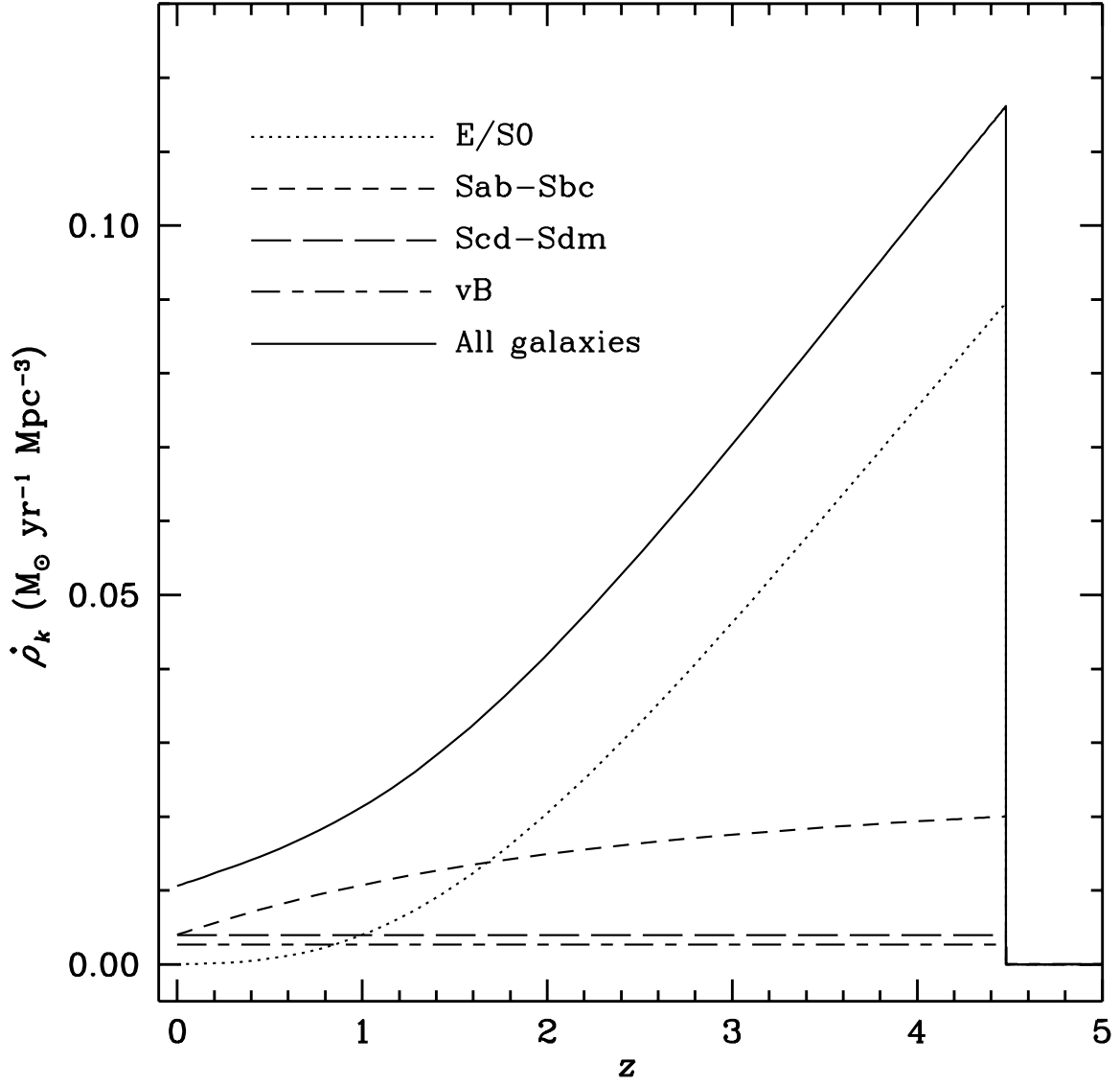}
\figcaption[Fig3.eps]{Evolution of the SFR density for the 4 galaxy
types, derived using the parametric SFRs and luminosity functions of
Pozzetti \etal\ The solid curve represents the global SFR density. 
\label{fig:sfr}}
\end{figure}

\begin{figure}
\plotone{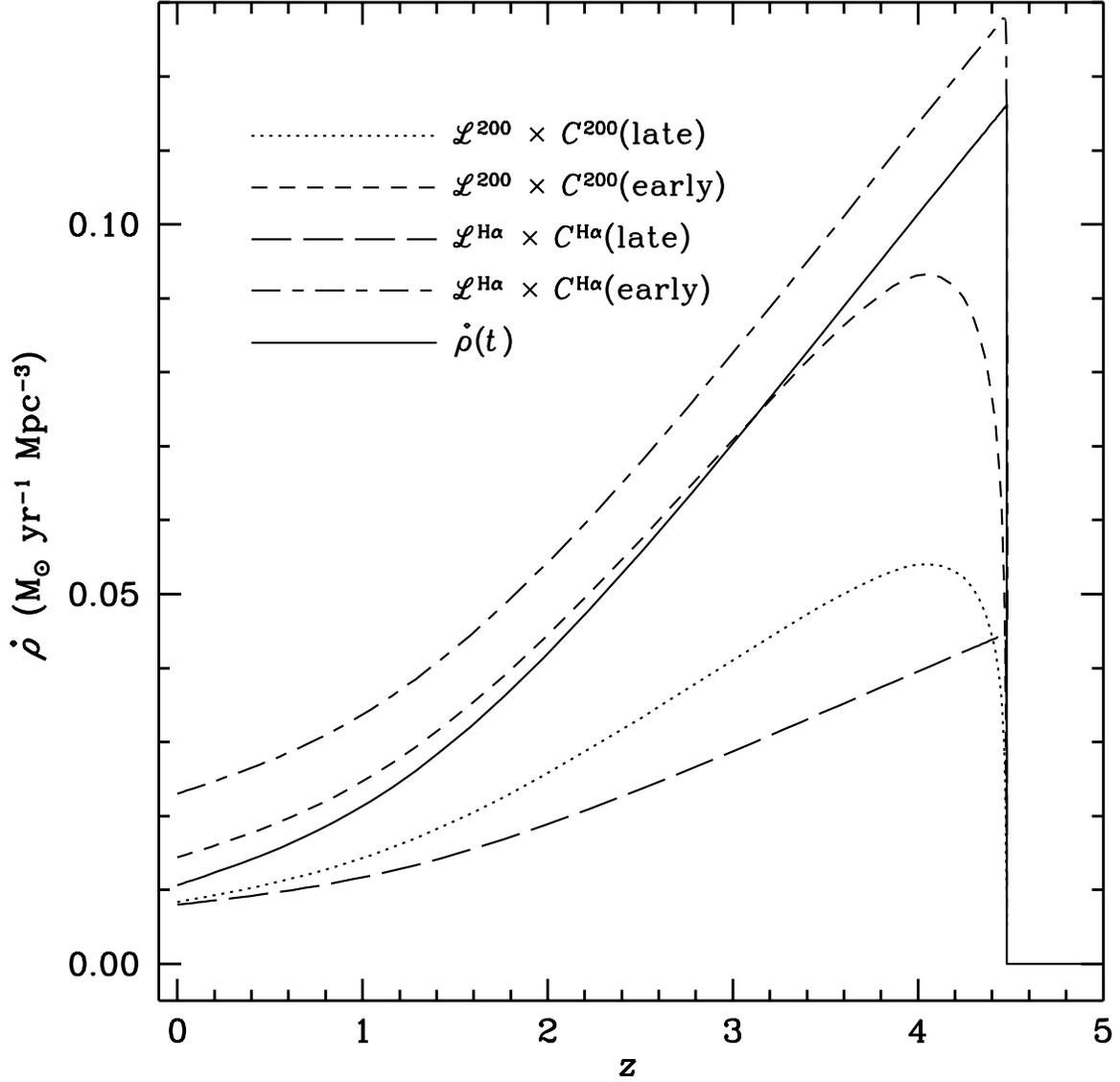}
\figcaption[Fig4.eps]{Evolution of the estimated global SFR density,
using the calibrations derived as explained in the text ($C^p$) and
listed in Table~\ref{tab:cs}. There are calibrations for both Salpeter
(late) and Scalo (early) IMFs. The solid curve is the actual global
SFR density transferred from Fig. 3. \label{fig:sfrcal}}
\end{figure}

Figure~\ref{fig:sfr} exhibits the time dependent star formation rate
density of each galaxy type, and for the totality of the
populations. Early-type galaxies dominate star formation for $z > 1$,
while all types except E/S0 contribute for $z \approx
0$. Figure~\ref{fig:sfrcal} shows the global star formation history
that would be inferred from the luminosity densities using each of the
calibration factors listed in Table~\ref{tab:cs}. Despite the
self-consistency in our approach, in no case does the inferred value
match the actual star formation history shown as the solid line in
Figures~\ref{fig:sfr} and \ref{fig:sfrcal}.

There are two reasons for the discrepancies seen in
Figure~\ref{fig:sfrcal} at $z=0$. First, the fact that the early and
late-type populations have different IMFs implies that neither a
Salpeter nor a Scalo calibration factor applied to the global
luminosity density will yield the true star formation rate. Secondly,
even in the absence of this difference, the 200~nm and \Halpha\
calibrations are not consistent because they refer to different
averages over the recent star formation history. This can be seen
clearly in Figure~\ref{fig:cevol}, where the vB and Scd-Sdm components
have identical values of $R_k^{{\rm H}\alpha}(t)$ after the initial
transient phase of $\approx 10$ Myr, but have different values of
$R_k^{200}(t)$ except at 100 Myr.

Another way to view the discrepancy is to contrast the relative
contribution of each galaxy type to the luminosity density with its
contribution to the star formation rate. Table~\ref{tab:ld} shows, 
for example, that at $z=0$ the vB galaxy type contributes 32\% 
to ${\cal L}^{{\rm H}\alpha}$ and 27\% to ${\cal L}^{200}$, while 
its contribution to the star formation rate itself is 24\%.

At high redshift yet another factor comes into play: the relative mix
of galaxy types changes as each undergoes luminosity evolution with
its specified star formation history. The systematic errors seen at
$z=0$ therefore change with redshift. Not surprisingly, we see that
for each waveband the Scalo IMF calibration is poorer than the
Salpeter at $z=0$, and that the situation reverses at $z=2$, since the
early-type galaxies become dominant at higher redshifts. There are,
however, always systematic errors in the inferred star formation
rates.

\section{DISCUSSION AND PROSPECTS}

\begin{figure}
\plotone{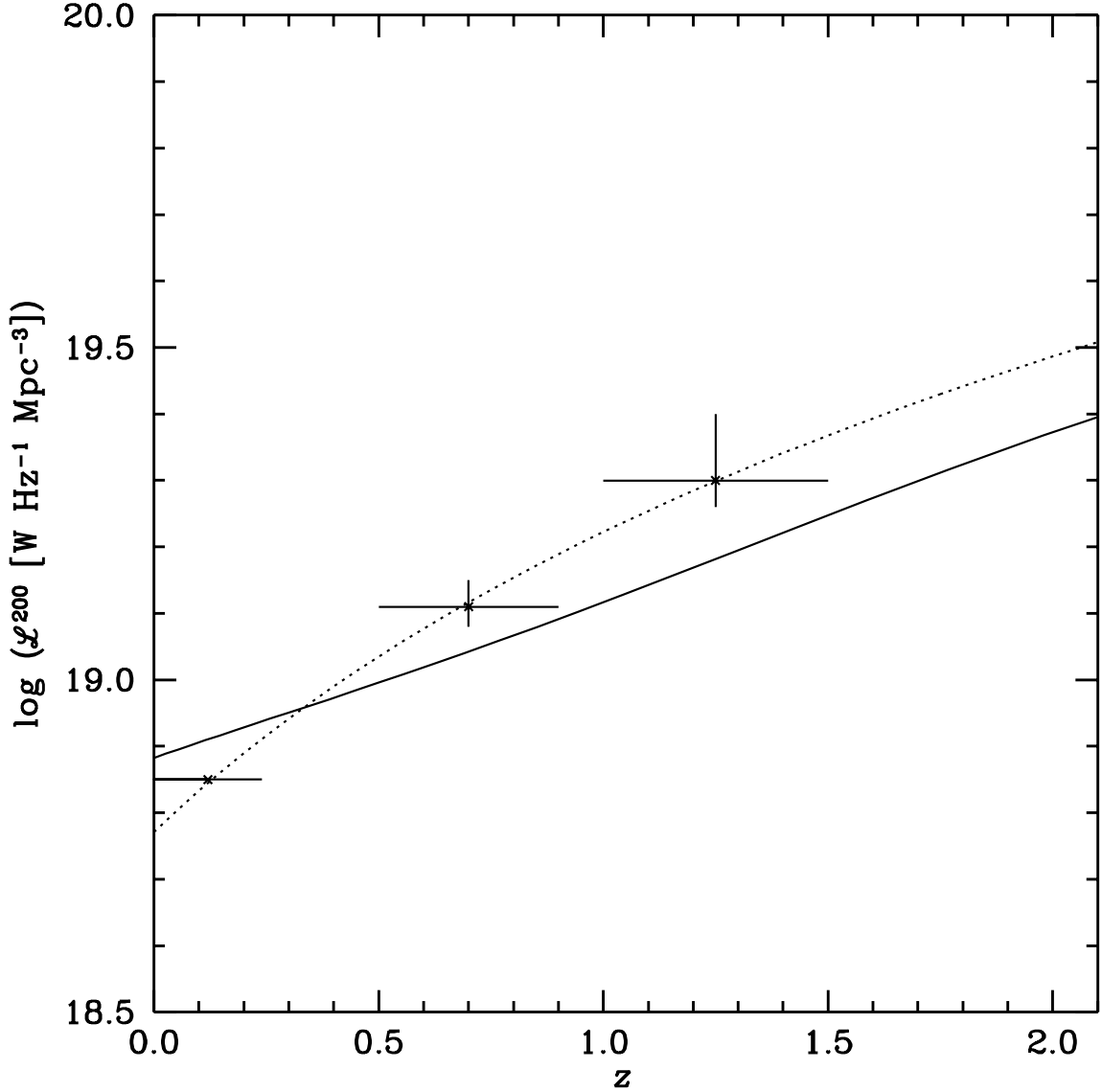}
\figcaption[Fig1.eps]{Rest frame 200~nm luminosity density predicted
using the model universe (solid line), compared with the values
measured by \citet{Cow99} (crosses) and fitted to a parametric form
(dotted line). The measurements adopt
$H_0=65$~km~s$^{-1}$~Mpc$^{-1}$. Conversion to the cosmology used in
this paper would reduce the measured luminosity density by $\approx
0.4$ dex but a correction for UV extinction would increase the value
by a similar amount. \label{fig:ldens}}
\end{figure}

The qualitative trends in our results could have been anticipated on
the basis of previous studies of the influence of the IMF and the star
formation history on the calibration of luminosity densities in terms
of star formation rates \citep[e.g.][]{Sch99}. Our study shows
quantitatively that attempts to infer the SFR density locally and over
a range of redshifts can be compromised by the presence of different
galaxy types, whose mix evolves differentially. Such differential
evolution is an almost inevitable consequence of models based on pure
luminosity evolution or, indeed, other descriptions of cosmic
evolution. Insofar as the model of \citet{Poz96} is typical in respect
of its mixture of galaxy types, systematic errors can be anticipated
of the order of a factor of at least 2 in both the absolute value of
SFR and in its relative evolution in $0 < z < 2$.  Some of the
intrinsic problems that arise from the adoption of fixed calibration
factors for the relation between SFR density and luminosity density
can be partially addressed by computing the luminosity density
explicitly from a model universe for comparison with observations
(cf. Figure~\ref{fig:ldens}).

At first sight, the inconsistency between the SFR inferred from 200~nm
and \Halpha\ calibrations could be regarded as a serious
problem. However, the difference between the two values contains
potentially useful diagnostic power regarding the star formation
history. Observations of a variety of different diagnostics of the
star formation rate in a sample of galaxies could provide a more
robustly constrained star formation history, by allowing the
determination of other important factors (such as type-specific
IMFs). However, the number of star formation diagnostics accessible to
observation is not large, and there are many parameters to be
constrained. Clearly, the possible existence of type-specific IMFs and
star-formation histories presents a significant challenge to
any systematic investigation of the cosmic evolution of
star formation.

\acknowledgments

We wish to thank Michael Rowan-Robinson for helpful comments and suggestions. 
JA gratefully acknowledges support in the form of a scholarship from the 
Science and Technology Foundation (FCT, Portugal) through Program Praxis XXI.

\clearpage

\begin{deluxetable}{ccc}
\tablewidth{0pc}
\tablecaption{Calibration constants $C^p$ \label{tab:cs}}
\tablehead{
\colhead{Galaxy type} & \colhead{$C^{200}$} & \colhead{$C^{{\rm H}\alpha}$}\\
\colhead{} & \colhead{M$_{\odot}$ yr$^{-1}$ / W Hz$^{-1}$} & 
\colhead{M$_{\odot}$ yr$^{-1}$ / W}}
\startdata
early &  $1.887 \times 10^{-21}$ & $3.381 \times 10^{-34}$ \\
late & $1.094 \times 10^{-21}$ & $1.176 \times 10^{-34}$ \\
\enddata
\end{deluxetable}

\clearpage

\begin{deluxetable}{lcrcrcr}
\tablecolumns{7}
\tablewidth{0pc}
\tablecaption{Type-specific contributions to luminosity and SFR densities 
at $z=0$ \label{tab:ld}}
\tablehead{
\colhead{} & \colhead{} & \colhead{} & \multicolumn{2}{c}{200~nm} & 
\multicolumn{2}{c}{\Halpha}\\
\colhead{Type}& \colhead{$\dot{\rho}_{k}$} & \colhead{\%
} & \colhead{log (${\cal L}^{200}$)}& \colhead{\%
} & \colhead{log (${\cal L}^{{\rm H}\alpha}$)} & \colhead{\%
}}
\startdata
E/S0   & 0.0000 & 0 & 16.58 &  0 & 28.95 &  0 \\ 
Sab-cd & 0.0040 & 38 & 18.31 & 27 & 31.08 & 18 \\ 
Scd-dm & 0.0040 & 38 & 18.55 & 46 & 31.53 & 50 \\ 
vB     & 0.0026 & 24 & 18.31 & 27 & 31.34 & 32 \\ 
Total  & 0.0106 & 100 & 18.88 & 100 & 31.83 & 100 \\
\enddata
\tablecomments{$\dot{\rho}_{k}$ in M$_{\odot}$ yr$^{-1}$ Mpc$^{-3}$, 
${\cal L}^{200}$ in W Hz$^{-1}$ Mpc$^{-3}$ and ${\cal L}^{{\rm H}\alpha}$ 
in W Mpc$^{-3}$}
\end{deluxetable}

\clearpage

\clearpage

\clearpage

\clearpage

\end{document}